# Evolution and Control of Oxygen Order in a Cuprate Superconductor

Nicola Poccia[1], Michela Fratini[1], Alessandro Ricci[1], Gaetano Campi[2], Luisa Barba[3], Alessandra Vittorini-Orgeas[1], Ginestra Bianconi[4], Gabriel Aeppli[5], Antonio Bianconi[1]

[1]Department of Physics, Sapienza University of Rome, P. le A. Moro 2, 00185 Rome, Italy

[2]Institute of Crystallography, CNR, via Salaria Km 29.300, Monterotondo Stazione, Roma, I-00016, Italy

[3]Elettra Sincrotrone Trieste. Strada Statale 14 - km 163,5, AREA Science Park, 34149 Basovizza, Trieste, Italy

[4] Department of Physics, Northeastern University, 360 Huntington Ave., Boston, Massachusetts 02115, USA.

[5]London Centre for Nanotechnology and Department of Physics and Astronomy, University College London, 17-19 Gordon Street, London WC1H 0AH, UK

[+]E-mail: antonio.bianconi@uniroma1.it

The disposition of defects in metal oxides is a key attribute exploited for applications from fuel cells and catalysts to superconducting devices and memristors. The most typical defects are mobile excess oxygens and oxygen vacancies, and can be manipulated by a variety of thermal protocols as well as optical and dc electric fields. Here we report the X-ray writing of high-quality superconducting regions, derived from defect ordering[1], in the superoxygenated layered cuprate, $La_2CuO_{4+y}$. Irradiation of a poor superconductor prepared by rapid thermal quenching results first in growth of ordered regions, with an enhancement of superconductivity becoming visible only after a waiting time, as is characteristic of other systems such as ferroelectrics[2,3] where strain must be accommodated for order to become extended. However, in $La_2CuO_{4+y}$, we are able to resolve all aspects of the growth of (oxygen) intercalant order, including an extraordinary excursion from low to high and back to low anisotropy of the ordered regions. We can also clearly associate the onset of high quality superconductivity with defect ordering in two dimensions. Additional experiments with small beams demonstrate a photoresist-free, single-step strategy for writing functional materials.

Illumination-induced structural transitions[4-7] are key phenomena in all of science, from photosynthesis in biology to photodoping[8] in semiconductor physics. Transition metal oxides, including both superconducting cuprates[9-15] and manganites[16] with their multiplicity of charge, orbital and spin ordered states[17,18], also exhibit such transitions. In particular, photo-switching occurs in $YBa_2Cu_3O_{6+y}$ and $La_2CuO_{4+y}$, with mobile oxygen ions, but the microscopic mechanism is not clear[9-15]. Although there have been several efforts to study photoconductivity and its microscopic origin in cuprates, direct structural measurements of the photo-induced redistribution of mobile oxygen interstitials (i-O) are lacking. Our present X-ray measurements fill this experimental gap, allowing us to correlate nano- and microscale structural configurations with material functionality (superconductivity) and to control these configurations to create new microstructures.

$La_2CuO_{4+y}$ is converted from a Mott insulator to a high temperature superconductor (HTS) by inserting i-O into the rocksalt $La_2O_2^R$ layers intercalated between the superconducting $CuO_2$ planes forming both random glassy and crystalline Q2 phases.[1] At optimum doping after long annealing, the system reaches a stationary state of scale-free organization of the Q2 grains. The i-O are mobile because of a 4% tensile lattice microstrain in the spacer $La_2O_2^R$ layers due to its natural misfit with the superconducting $CuO_2$ planes[19,20]. Even though they are responsible for the appearance of high quality superconductivity, the dynamics of i-O ordering are unknown. Understanding these domain growth dynamics[21,22] will allow the definition of photopatterning protocols for $La_2CuO_{4+y}$ itself as well as providing the basic principles for the organization of oxygen defects in layered transition metal oxides.

Because of their obvious relevance to photolithography, we focus here on persistent effects, meaning those, such as an increase of $T_c$, which survive after photostimulation ends. First, we show that the persistent effects in $La_2CuO_{4+y}$ single crystals are due to photo-stimulated i-O mobility, and second, we give a detailed quantitative description of the photoinduced non-equilibrium disorder-to-order transition of i-O in the spacer layers. For our X-ray annealing experiments, the sample is maintained at temperatures $T$ with $220<T<300K$[1,23,24] so that the full growth and coarsening process can be observed within a reasonable measuring time (approximately 15 h); in the dark or at lower $T$, i-O ordering takes place on timescales of weeks and months, meaning that it could not been sensibly studied. The disordered i-O phase was obtained by quenching the sample from 370K (above the order-to-disorder transition critical temperature 330K) to below 300 K. We measured the disorder-to-order transition of i-O domains by a novel approach using x-ray continuous flux for photo-switching the transition at the Trieste synchrotron radiation facility Elettra. The technique allows structural investigation by x-ray diffraction (XRD) of the same portion of the sample illuminated by x-rays as shown in Fig. 1.

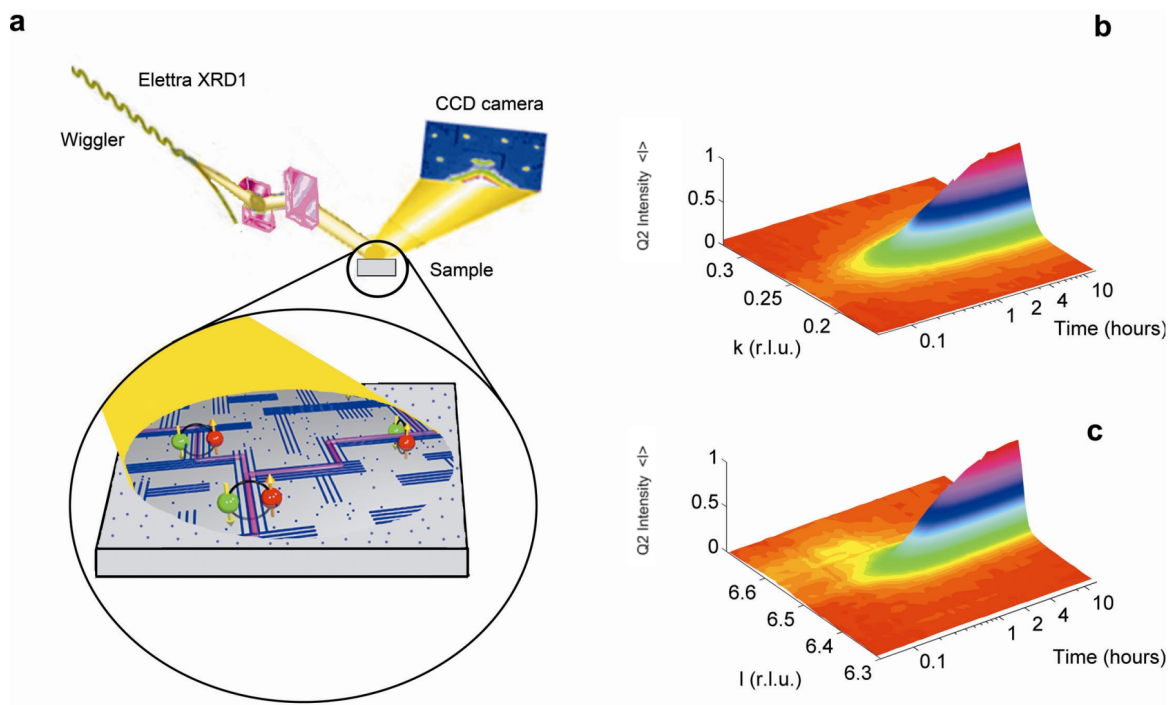

Figure 1| **X-ray photoinduced i-O ordering experiment. a**. The x-ray beam emitted by the wiggler source is focused on the sample surface at the X-ray diffraction beamline (XRD1) at Elettra (Trieste) synchrotron radiation facility. A charge-coupled area detector (CCD) records the X-ray reflections from the illuminated sample area. surface layer of about 1.5 micron thickness. The striped Q2 domains are photo-switched in the same surface layer thickness of the sample. **b** and **c** panels show the Q2 peak profiles along the in plane k-direction (panel b) and the out-of-plane l-direction (panel c) as a function of time of x-ray exposure at constant x-ray photon flux $\Phi_{P(0.1nm)} = 5 \cdot 10^{14} N_{P(0.1nm)} \cdot s^{-1} \cdot cm^{-2}$ and at fixed temperature 250 K, starting from the quenched disordered i-O phase with suppressed Q2 satellite reflections. There is clear evidence for both a time threshold before i-O ordering process starts and also for a lack of saturation of the intensity growth after 14 hours.

We establish (Fig. 1a) the photo-induced structural effects in the illuminated area by measuring the time evolution of the XRD pattern recorded by a CCD area detector. The reciprocal-space image reveals the satellites, disappearing above the order-to-disorder transition temperature for i-O at 330K, and associated with the three-dimensional Q2 superstructure of the i-O dopants, displaced from the main crystal reflections (004) (006) (008) of the orthorhombic Fmmm crystal structure by the wave-vector $\mathbf{q}_2$(0.09a*,0.25 b*,0.5c*). We monitor the i-O ordering by recording the Q2 satellite peaks as a function of time *t* after rapidly quenching the sample from the high temperature disordered phase at 370 K. Figs. 1b,c shows our key experimental discovery, namely that for a constant photon flux, the intensity of the Q2 x-ray satellite reflections undergoes a non-linear rise with *t*. Therefore, as for $YBa_2Cu_3O_{6+y}$[14-16], there is an increase of i-O mobility in $La_2CuO_{4+y}$, induced by the continuous x-ray photon flux increasing the disorder-to-order conversion rate of the non-equilibrium annealing process by at least two orders of magnitude. Furthermore, the satellite reflections become prominent only after an incubation time $t_0$ which is a clear indication of the nucleation and growth phenomena typical for electrically switched[0,0] and photo-induced disorder-to-order phase transitions[4-6] seen for heterostructures at atomic limit, such as the cuprates.[19]

Nucleation and growth phenomena have been much studied in three-dimensional systems. $La_2CuO_{4+y}$ offers an unprecedented host for their observation, in a layered system because all characteristics – notably widths and intensities - of the Q2 peaks are easily determined using synchrotron X-ray diffraction. Figure 2 shows the outcome of quantitative analysis, to establish the integrated Q2 intensities <I> (frame a), which measure the total volume of Q2 ordered regions, and their widths, inversely proportional to their typical extents (frame b) in all three spatial directions. Fig. 2a shows that domain growth dynamics has a threshold time before which appreciable change is visible in the integrated intensity; as detailed in the SI, the data deviate from the standard Kolmogorov formula $<I>(t) = (1 - e^{-(t/\tau)^4})$ for nucleation and growth. Fig. 2c shows how the superconducting $T_c$ evolves with the same photo-assisted anneal, allowing correlation with the status of the nucleation and growth. Q2 order and optimal superconductivity appear via a multistage process: (1) At t=0, there are very small and quite isotropic regions of Q2 order, extending roughly $a/(\pi \cdot FWHM) \approx 2.5 nm$ along both in-plane directions (*a* axis=0.536nm, *b* axis=0.542nm) and 4nm along the c-axis (1.313nm). From t=0 to just before the waiting time $t_0$ the Q2 regions grow visibly only in the in-plane direction, and there is a negligible increase in the overall volume occupied. (2) Just at $t_o$, there is a very rapid rise in <I> even while the dimensions of the clusters continue to grow smoothly and relatively slowly.

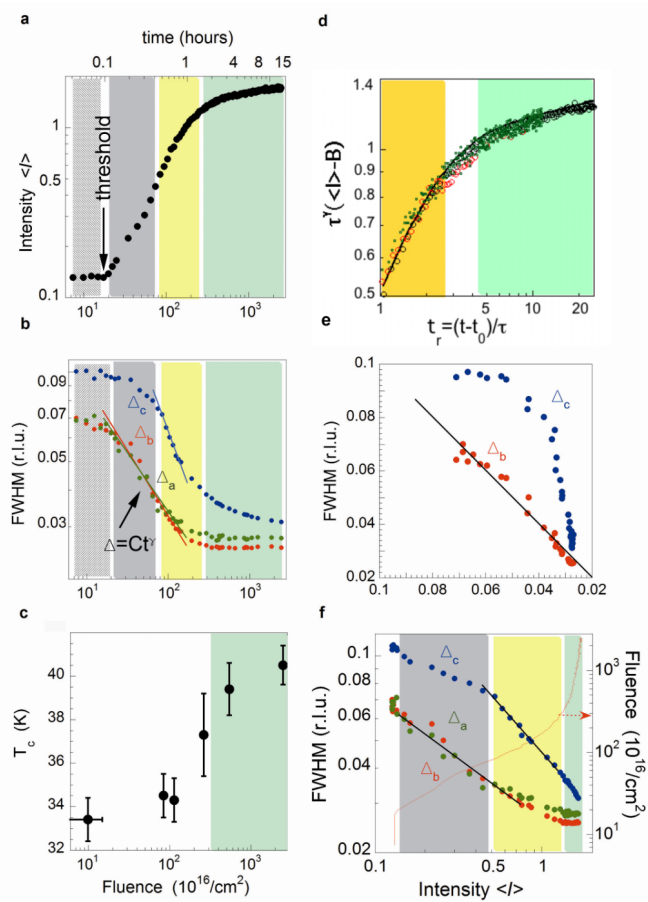

Figure 2| **The non linear growth dynamics**. | **a**. The integrated intensity of the Q2 XRD reflections is plotted as function of fluence $\phi$ and of time t. After an incubation period, $t < t_0$ we first see a nucleation regime (grey) where the XRD intensity increases rapidly with the growth of two-dimensional in-plane i-O ordered domains followed by the growth of a three dimensional order with a slow (yellow) and very slow (green) growth, lasting for at least 15 h of exposure| **b**. The full-width half-maximum (FWHM) in reciprocal lattice units (r.l.u.) of the Q2 satellite reflections along the a-axis, b-axis and c-axis respectively, $\Delta_a$, $\Delta_b$, and $\Delta_c$ are plotted. In the green regime the FWHM along a and b axis saturate while in the c axis direction continues to decrease. | **c**. The superconducting critical temperature $T_c$ of quenched samples at different point of the annealing process. $T_c$. The error bars indicate the divergence between the $T_c$ onset and the maximum of the derivative of surface resistivity. | **d**. The time evolution of the Q2 XRD intensity in the slow 3D growth regime can be fitted by a single curve that allows the scaling of different experiments run at different temperatures in the range 250-300K and different photon flux. All experimental curves, with different threshold $t_0$ and $\tau$, collapse on the same curve $\tau^\gamma (<I(t)> - B) \propto \left(1 - e^{-t_r}\right) \cdot (t_r)^\gamma$ where $t_r = (t - t_0)/\tau$, $\tau$ is a characteristic time constant at the onset of the 3D ordering regime (yellow in Fig. 2a) and B is the background at t=0. | **e**. The evolution of the FWHM of the profile of x-ray diffraction satellites along the b and c direction $\Delta b$, $\Delta c$ of the Q2 superstructure around the (006) main reflection as a function of the FWHM, $\Delta a$, of the profile of satellite reflection in the a direction. | **f**. The evolution of the FWHM of the profile of x-ray diffraction satellites along the a, b and c direction $\Delta a$, $\Delta b$, $\Delta c$ of the Q2 superstructure around the (006) main reflection as a function of the Intensity of the superstructure reflections.

Therefore, at this point we are passing from the conventional coarsening regime,[21] dominated by short range interactions within the two-dimensional planes, into a rapid growth phase (grey) where long range strain fields generated by the clusters bias the system towards the growth of more nuclei[0,0]. (3) After the sudden strain-induced increase in cluster growth, they begin to grow perpendicular to the planes while they continue to accrete within the planes. Well within this (yellow) regime, they actually start to grow faster along c than along a and b. Finally, (4), a regime (green) is entered where the growth is almost exclusively along the *c* axis, resulting, in the long-time limit in clusters with nearly the same anisotropy as we began with at t=0, but with spatial extents of approximately three times larger along each direction. It is once the in-plane growth has completed, but before the clusters have reached their full extent along *c*, that the superconducting $T_c$ undergoes its most pronounced rise.[23,24] The implication is that the optimal superconductivity is associated with i-O ordering induced Fermi surface reconstruction and electron hopping between layers[25].

The images in the right hand column of Fig. 2 show several key features of the Q2 nucleation and growth process. The first (Fig. 2a) concerns its defining functional form and universality. Here, we have carried out the experiment of Fig. 1 for several photon beam fluxes and constant sample temperatures $T_s$ between 220 K and 300 K, and consistently find the same waiting time phenomenon, characterized by the delay time $t_o$ which is simply proportional to the flux and approaches zero (within our measurement capability) as $T_s$ approaches 330 K.

Figure 2d shows that domain growth data in the growth regimes (3) and (4), collected at different temperatures and with different photon fluxes, all collapse onto the same curve. The integrated intensity follows the simple functional form $<I(t)> - B = C \cdot \left(1 - e^{-(t-t_o)/\tau}\right) \cdot \left(t - t_o\right)^y$ where $t = F_{P(0.1\ nm)}/\Phi_p$ and B is the intensity at t=0. Here τ and $t_0$, which scale with $t_0 = \frac{1}{3} e^{\tau/300}$, characterize the domain growth in the regime in the *c* axis direction.

One of our most remarkable results is the evolution of the anisotropy of the Q2 clusters, where we proceed from small and essentially isotropic clusters at t=0 to large and essentially isotropic clusters as t→∞, via intermediate states characterized by much higher anisotropy. The growth process follows Ostwald's rule of stage states[26] where the parent phase will first transform to a metastable phase (with more oblate Q2 clusters) closest to it in free energy, as for transitions between amorphous and crystalline phase[27] and in protein crystal growth[28]. Figure 2e shows this excursion where we treat the in-plane width $\Delta_a$ as the control parameter. The independently measured in-plane width $\Delta_b$ is indistinguishable from $\Delta_a$, and therefore falls on the diagonal, whereas $\Delta_c$ approaches the diagonal along horizontal and vertical straight lines for the small and large cluster limits corresponding to large and small $\Delta_a$ respectively.

An important problem in the non-equilibrium physics of first order phase transitions is to establish the relative importance of nucleation and growth.[3] A particularly simple analysis tool is a plot of the widths, corresponding to the sizes of the ordered clusters against the integrated intensity <I>, reflecting the total volume occupied by the ordered phase. Figure 2f is such a plot for our data and gives a revealing view of the different growth regimes. In particular, we see that the Q2 peak widths in the white regime before the threshold time $t_o$ are actually growing much faster than *I*, implying that, initially different ordered regions are simply combining, leaving the net Q2 volume essentially unchanged. Immediately after $t_o$, the Q2 widths follow a power law $\Delta = I^{-\alpha}$ with exponents $\alpha_a \approx \alpha_b \approx 0.39 \pm 0.02$ and $\alpha_{c(I)} = 0.2 \pm 0.02$. The in-plane exponents are difficult to distinguish, but the c-axis exponent is much smaller. Furthermore, the sum of the three exponents is indistinguishable from unity, indicating that what controls the rising volume fraction of Q2 ordered regions is the in-plane growth of pre-existing clusters. On the other hand, for times well beyond $t_o$, we enter the yellow regime where $\alpha_a$ and $\alpha_b$ remain unchanged but $\alpha_{c(II)} = 0.65 \pm 0.02$ is very much higher. The exponents now sum to appreciably more than unity, meaning that this is a growth and agglomeration regime, where clusters are merging (particularly along c) as well as growing, thus adding total volume fraction of Q2 ordered regions at a smaller rate than in a pure, independent cluster, growth regime. In the final growth regime (green), the size of the Q2 grains have reaches a maximum size and the reflection profile in the *b* and *a* directions reaches saturation while the FWHM of reflections in the *c* direction decreases with a power law exponent $\alpha_{c(III)} = 0.89 \pm 0.15$, an exponent close to unity implying growth of the ordered phase mainly via further growth of existing clusters along the *c* axis.

Although each regime in Fig. 2f is clearly unique, it is worth repeating that a common feature is that the sum $\alpha_a+\alpha_b+\alpha_c$ of the exponents is never less than unity (within error), implying that there is never any need for new nuclei for Q2 ordered regions beyond those which exist after the quench. The entire process by which <I> increases is accounted for by growth and agglomeration of preexisting nuclei.

Having extensively characterized the non-equilibrium disorder-to-order process induced by X-ray irradiation, we show here that the observed phenomena can be exploited for writing and reading Q2 patterns, leading to possible technological applications of writing and erasing planar superconducting electronic devices.

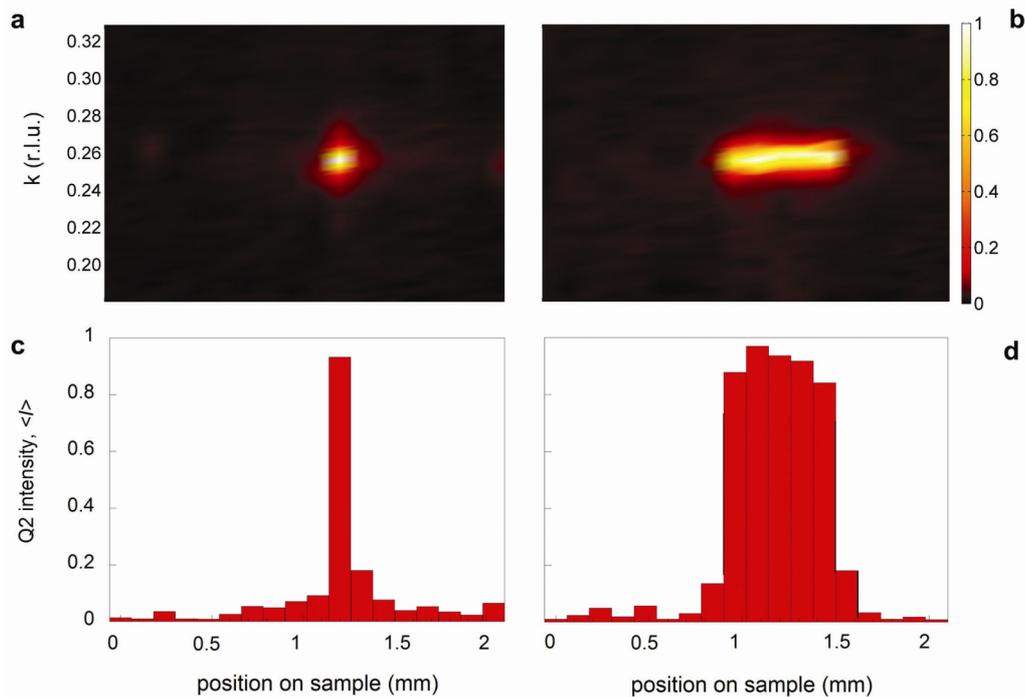

Figure 3| **X-ray manipulation of high-$T_c$ superconductors. a,b** The formation of a dot and of a stripe of Q2 i-O fractal phase embedded in the disordered phase. A surface spot has been illuminated by the X-ray beam well for a time larger than $t_0 + 2\tau$ with a x-ray fluence well above threshold keeping constant temperature at 250K. Therefore the fractal i-O phase is formed in the illuminated portion of the sample surface, as shown in Fig. 1, while in the masked part the i-O are in the disordered phase. After illumination for writing we have read the sample surface using a low x-ray flux beam, below threshold for non invasive x-ray diffraction. The lower figures show 3D color plots of the XRD Q2 satellite reflection. **c,d** The integrated intensity is plotted as a function of the position of the beam on sample showing the formation of the 100 μm dot and of the stripe. The dot and the stripe can be easily erased by local increase of the sample temperature above 350 K. This technique allows switching (on-off) the self-organized microstructures on a nanometer scale and to control them in situ by synchrotron x-ray scattering.

We have followed a multistep protocol starting with warming the sample to 380 K, where the Q2 peaks disappear entirely (erasing). The sample is then quenched down to 200-300 K and illuminated at a constant temperature. At this point it is possible to write patterns like the dot in Fig. 3a or the stripe in Fig 3b using a photon beam size of 100 μm with flux of $10^{15}$ photons/cm$^2$. In fact, under continuous illumination the Q2 superstructure peak is formed due to the photo-conversion of i-O disordered phase to the i-O scale-free ordered phase. These results clearly show that it is possible to read and write microstructures, with edges defined to within at least the 100 μm step size in our current demonstration experiments, formed by two different superconducting materials, namely the first (black) with disordered i-O and the second (red) with scale-free ordered i-O in the spacer layers as shown in Fig. 3c,d.

As $t \to \infty$ the FWHM of the i-O peaks remains at the relatively high value of 0.025 r.l.u., which imposes the very small lower bound of $a/(\pi \cdot FWHM) \approx 6 \; nm$ on the possible resolution of future nanolithography employing advanced, highly focused X-ray beams. It is also worth noting that the X-rays can be used in combination with photons of other wavelengths to cause disorder-to-order phase transitions in the recording material, in analogy with optical CD data recording. In particular, for erasure, the necessary increase of the sample temperature above the disorder-to-order transition can be achieved in situ via higher flux laser illumination or microwave beam irradiation. The lowest level of X-ray photon flux (below the threshold) can be used for checking the Q2 pattern without damaging the superconductor surface. An initial device could be manufactured by simply drawing a wire of Q2 domains across the sample via continuous translation of the sample stage under an X-ray beam defined by modern focusing optics exploiting a Fresnel zone plate. In the middle of the wire drawing process, the fluence is reduced below its critical value for Q2 formation via either a brief acceleration of the stage or a shutter closing off the beam, leaving behind the desired weak link. The stage can be manipulated to produce more complex circuits, including wire loops containing weak links, and the stability, for at least weeks, of the disordered phase at room temperature in the absence of radiation assures the possibility of moving a prepared circuit from the lithography station to a electrical measuring cryostat away from the X-ray source. We note also the possibility of patterning not only with X-rays, but also with electron beam writers, which are much more common and routinely produce smaller beams.

Given that electronic inhomogeneities are the prerequisite for interesting multi-terminal devices such as transistors and tunnel junctions, our results open the way to novel superconducting electronics and their manipulation. Moreover the control and manipulation via x-ray excitation of oxygen interstitials or vacancies organization can also be used for modulating complexity and writing devices in manganites[5,16-18] and at the interfaces of functional oxide superlattices.

**Methods**

The X-ray diffraction experiment uses a photon beam of monochromatic, 0.1 nm wavelength (12.4 KeV) x-rays emitted from the 2 GeV electron storage ring Elettra at Trieste. The X-ray beam spot on the sample surface has dimensions in the range between $400 \times 400$ to $100 \times 100$ μm². The intensity, I(Q2), of the superstructure satellites due to the Q2 ordering of (i-O) in the $La_2CuO_{4.1}$ crystal is integrated over square sub-areas of the images recorded by the CCD detector and then normalized to the intensity ($I_0$) of the tail of the main crystalline reflections. The synchrotron X-ray photons, are focused on a spot of 100-400 microns on the sample surface with a flux (i.e., the number of photons shining the sample surface per second per unit area) of $\Phi_{P(0.1\ nm)} = f \cdot 10^{14} N_{P(0.1\ nm)} \cdot s^{-1} \cdot cm^{-2}$ where $f$ can be controlled and varied in the range $1 < f < 10$. For the typical case where $f = 5$ the power density on the sample surface is 1 Watt/cm², equivalent to an optical (510 nm) laser flux of $\Phi_{P(510\ nm)} = 2.5 \cdot 10^{18} N_P \cdot s^{-1} \cdot cm^{-2}$ as shown in Fig. 1. The x-ray penetration depth is about 1.5 μm, implying an illuminated areal density of Cu ions of the order of $10^{18}\ cm^{-2}$. Considering that an energy of 2 eV is required to create an electron-hole pair in cuprates, this X-ray flux can induce photo-induced persistent effects in a micron thick layer much like those which visible lasers induce in much thinner thin films.[11-16] The effect of continuous illumination corresponds to photo-excitation in which the state-changing rate is proportional to the intensity of light. Therefore the physical state of the system is controlled by the fluence $F_{P(0.1\ nm)}(N_{P(0.1nm)} \cdot cm^{-2}) = \Phi_p \cdot t$; we have checked that on doubling or halving the flux, the time scales are respectively halved or doubled.

**Author Information** The authors declare no competing financial interests. Correspondence and requests for materials should be addressed to A.B. (e-mail: antonio.bianconi@uniroma1.it).

**Acknowledgements.** We thank S. Agrestini, M. Colapietro, D. Di Castro, the Elettra XRD beamline staff in Trieste for experimental help and R. Markiewicz and A. Coniglio for useful discussions. This experimental work has been carried out with the financial support of the European STREP project 517039 "Controlling Mesoscopic Phase Separation" (COMEPHS) (2005-2008) and Sapienza University of Rome, research project "Stripes and High-$T_c$ Superconductivity".

**Author Contributions** All authors contributed to providing experimental support, interpreting data and writing the manuscript in particular A.B., L.B., G.B., G.C., M.F., N.P., A.R. & A.V.-O. have performed the experiment, A.B., G.C., M.F., N.P., & A.R., have performed sample manipulations, L.B. provided the X-ray beamline, N.P., G.A., G.B. M.F. & A.V.-O the data analysis, G.B. provided theoretical support and G.A., A.B., & N.P., have written the paper.